\documentclass[prl,letterpaper,twocolumn,showpacs,superscriptaddress,amsmath,amsfonts,amssymb,preprintnumbers]{revtex4}

\usepackage[T1]{fontenc}\usepackage[latin1]{inputenc}
\usepackage{dcolumn,graphicx,color} 
\renewcommand{\figurename}{Fig.}
\makeatletter\renewcommand{\fnum@figure}[1]{\figurename~\thefigure~(color online).}\makeatother

\newcommand{\champagne}{\kern-1.1ex
\raisebox{-0.35pt}{
\includegraphics[width=1.34ex]{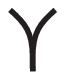}}
\kern-4.5pt}

\newcommand{\waterfalls}{\kern-1.1ex
\raisebox{-0.35pt}{
\includegraphics[width=1.34ex]{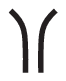}}
\kern-3.6pt}

\begin{document} \pagestyle{plain}

\title{Momentum and excitation energy dependence of the ``waterfalls'' in cuprates.}

\author{D.\,S.\,Inosov}
\affiliation{Institute for Solid State Research, IFW Dresden, P.\,O.\,Box 270116, D-01171 Dresden, Germany.}
\author{J.\,Fink}
\affiliation{BESSY GmbH, Albert-Einstein-Strasse 15, 12489 Berlin, Germany.}%
\affiliation{Institute for Solid State Research, IFW Dresden, P.\,O.\,Box 270116, D-01171 Dresden, Germany.}
\author{A.\,A.\,Kordyuk}
\affiliation{Institute for Solid State Research, IFW Dresden, P.\,O.\,Box 270116, D-01171 Dresden, Germany.}
\affiliation{Institute of Metal Physics of National Academy of Sciences of Ukraine, 03142 Kyiv, Ukraine.}
\author{S.\,V.~Borisenko}\author{V.\,B.\,Zabolotnyy}\author{R.~Schuster}\author{M.\,Knupfer}\author{B.\,Büchner}
\affiliation{Institute for Solid State Research, IFW Dresden, P.\,O.\,Box 270116, D-01171 Dresden, Germany.}
\author{R.\,Follath}\author{H.\,A.\,Dürr}\author{W.\,Eberhardt}
\affiliation{BESSY GmbH, Albert-Einstein-Strasse 15, 12489 Berlin, Germany.}
\author{V.~Hinkov}\author{B.\,Keimer}
\affiliation{Max-Planck-Institut for Solid State Research, 70569 Stuttgart, Germany}
\author{H.\,Berger}
\affiliation{Institut de Physique de la Mati\`{e}re Complexe, EPFL, 1015 Lausanne, Switzerland.}




\begin{abstract}
Using high-resolution angle-resolved photoemission spectroscopy we have studied the momentum and photon energy
dependence of the anomalous high-energy dispersion, termed ``waterfalls'', between the Fermi level and 1~eV binding
energy in several high-$T_\text{c}$ superconductors. We observe strong changes of the dispersion between different
Brillouin zones and a strong dependence on the photon energy around 75~eV, which we associate with the resonant
photoemission at the Cu\,3p\,$\rightarrow$\,3d$_{x^2-y^2}$ edge. We conclude that the high-energy ``waterfall''
dispersion results from a strong suppression of the photoemission intensity at the center of the Brillouin zone due to
matrix element effects and is, therefore, not an intrinsic feature of the spectral function. This indicates that the new
high energy scale in the electronic structure of cuprates derived from the ``waterfall''-like dispersion may be
incorrect.
\end{abstract}

\maketitle

\textit{Introduction}. It is widely believed that the study of many-body effects in the electronic structure of layered
cuprates is a possible clue to the mechanism of high-temperature superconductivity in these materials \cite{Reviews}.
However, even after years of intense studies there is still no full understanding of the renormalization effects and of
the relevant energy scales in their electronic excitation spectrum.

Appearance of the new generation of electron spectrometers with the wide acceptance angle ($\pm$\,15$^\circ$ for
\textit{Scienta~R4000}) has opened up the possibility of viewing the electronic structure of cuprates over a broad
momentum range covering more than one Brillouin zone (BZ) \cite{BZnote} in a single measurement. This has triggered a
series of publications evidencing anomalous high-energy dispersion in the renormalized band structure of
Bi$_2$Sr$_2$CaCu$_2$O$_{8+\delta}$ \cite{XieYangShen06, MeevasanaZhou06, GrafMcElroy06, GrafLanzara06, PanRichard06,
VallaKidd06}, Bi$_2$Sr$_2$CuO$_{6+\delta}$ \cite{MeevasanaZhou06, PanRichard06, GrafLanzara06}, La$_{2-x}$Sr$_x$CuO$_4$
\cite{MeevasanaZhou06, GrafLanzara06, ChangPailhes06}, La$_{2-x}$Ba$_x$CuO$_4$ \cite{VallaKidd06},
Pr$_{1-x}$LaCe$_x$CuO$_4$ \cite{PanRichard06}, Ca$_2$CuO$_2$Cl$_2$ \cite{RonningShen05}, and
Ba$_2$Ca$_3$Cu$_4$O$_8$(O$_\delta$F$_{1-\delta}$)$_2$ \cite{MeevasanaZhou06} at the binding energies higher than
$\sim$\,0.3\,--\,0.5~eV\,---\,a region that has previously been scarcely explored. All of these reports seem to agree on
the qualitative appearence of the spectra: (i) in the $(0,0)$\,--\,$(\pi,\pi)$ (nodal) direction the
``\textit{high-energy kink}\kern.8pt'' at $\sim$\,0.4~eV is followed by a nearly vertical dispersion
(``\textit{waterfall}\kern.8pt'') that ends up below 1\,eV with a barely detectable band bottom approaching that of the
bare band; (ii) the band bifurcates near the high-energy kink, forming another branch with a bottom at $\sim$\,0.5~eV
\cite{GrafMcElroy06,PanRichard06,MeevasanaZhou06}; (iii) as one moves away from the nodal direction, the
``\textit{vertical dispersion}\kern.4pt'' persists surprisingly up to the $(\pi,0)$ (antinodal) point, forming a
``\kern-.5pt\textit{diamond}\kern.8pt'' shape in momentum space at $\sim$\,0.5~eV \cite{GrafLanzara06, VallaKidd06,
ChangPailhes06}; (iv) no dependence on the doping concentration, momentum, and photon energy has been detected so far.

Unfortunately there is still no consensus on the physics behind these phenomena. In principle any strong coupling to a
bosonic mode would lead to the appearance of the incoherent spectral weight below the energy of the mode
\cite{Engelsberg63}, which can resemble the ``vertical dispersion'', so distinguishing between different mechanisms is
impossible without accurate quantitative comparison between theory and experiment. Up to now, several qualitative
explanations have been proposed for the high-energy anomaly, including a disintegration of the quasiparticles into a
spinon and holon branch \cite{GrafMcElroy06}, coherence-incoherence crossover \cite{PanRichard06, ChangPailhes06},
disorder-localized band-tailing \cite{AlexandrovReynolds07}, polarons \cite{MeevasanaZhou06}, familiar $t$-$J$ model
with \cite{Manousakis06} or without \cite{WangTan06} string excitations, as well as the self-energy approach with strong
local spin correlations \cite{XieYangShen06}, itinerant spin fluctuations \cite{VallaKidd06, MeevasanaZhou06,
MacridinScalapino07}, or quantum criticality \cite{ZhuVarma07}. The reported ``diamond''-like momentum distribution of
the ``waterfalls'' in \hbox{Bi-2212} with its sides pinned at $(\pm\pi/4,\pm\pi/4)$ around the BZ center could be a sign
of BZ folding due to some form of antiferromagnetism \cite{GrafMcElroy06, PanRichard06}. However, such picture is
violated in other cuprates and does not appear to be universal \cite{VallaKidd06}.

\begin{figure*}
\vspace{-1em}\includegraphics[width=\textwidth]{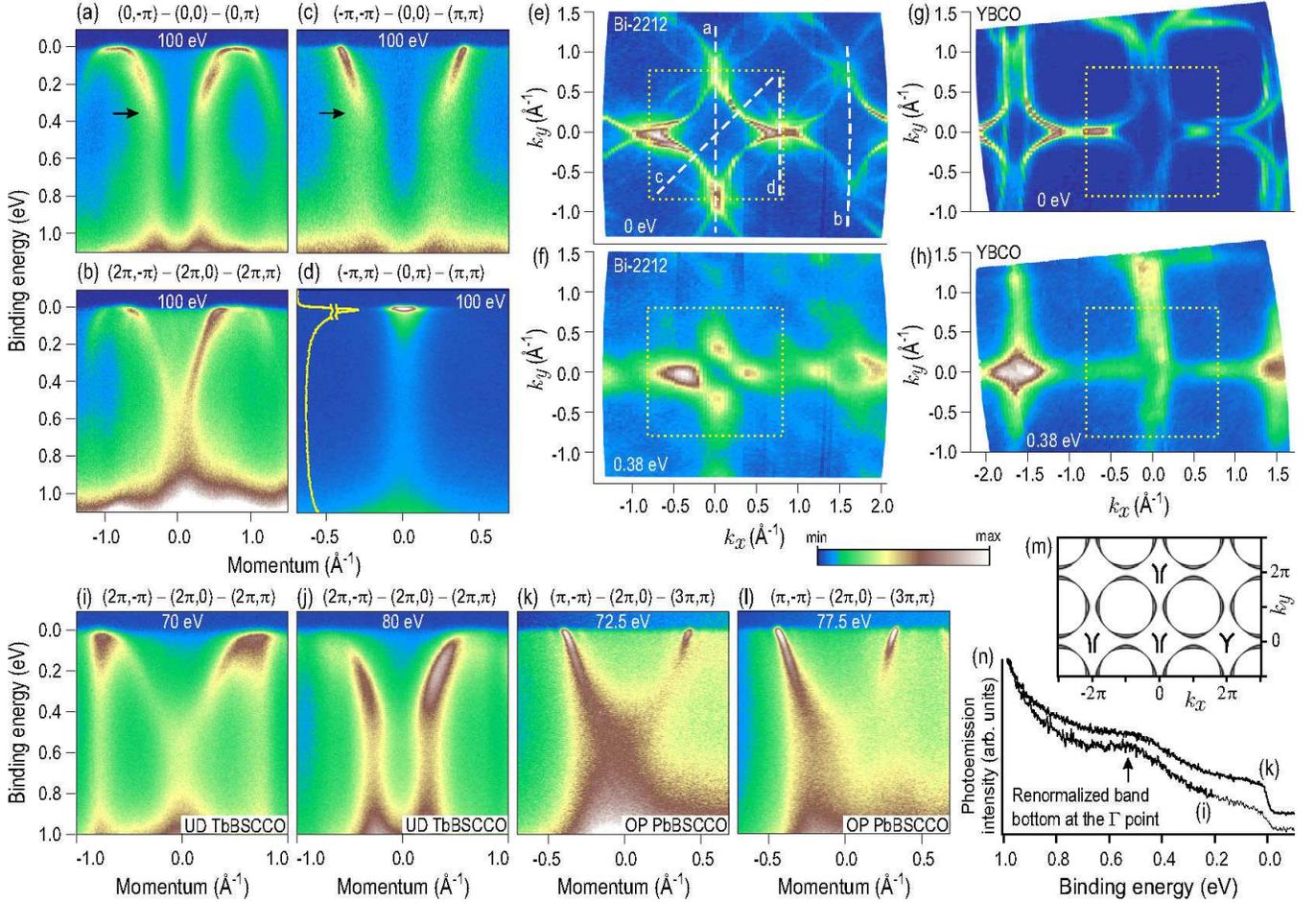}\caption{(\textbf{a})\,--\,(\textbf{h}) Typical snapshots of
the one-particle excitation spectra of Bi-2212 (\textbf{a}\,--\,\textbf{f}) and Y-123 (\textbf{g},\,\textbf{h}) measured
by angle-resolved photoemission with 100~eV photon energy. The spectra (\textbf{a})\,--\,(\textbf{d}) are measured along
the high-symmetry directions marked by the dashed lines on the Fermi surface map (\textbf{e}). Spectra (\textbf{a}) and
(\textbf{c}) from the 1st BZ exhibit strong high-energy kinks (black arrows) and ``waterfalls'', while the equivalent
spectrum (\textbf{b}) from the 2nd BZ exhibits no pronounced high-energy scales. Additional spectral weight is clearly
seen in panel (\textbf{d}), where the curve at the left of the panel shows the energy distribution curve at $(\pi,0)$.
(\textbf{f}) Constant-energy cut at 0.38\,eV below the Fermi level in Bi-2212 showing spectral weight depletion along
the 1st BZ diagonals. (\textbf{g}),\,(\textbf{h}) The respective constant-energy maps of Y-123. The 1st BZ on the
constant-energy maps is confined by the dotted squares. (\textbf{i}),\,(\textbf{j}) and (\textbf{k}),\,(\textbf{l})
Pairs of equivalent spectra of Tb- and Pb-doped Bi-2212 taken in the 2nd BZ along the $(\pi,0)$ and $(\pi,\pi)$
directions with two different excitation energies as indicated on top of each panel. In both directions, the onset of
the ``waterfalls'' behavior suddenly occurs at about 75 eV photon energy. The color scale in all panels represents
photoelectron intensity. The spectra are normalized to the background above the Fermi level. The spectra in panels
(\textbf{k}) and (\textbf{l}) are in addition multiplied by a linear function of momentum to enhance the right-hand part
of the spectrum, which otherwise has much lower intensity than the left-hand part due to the experimental geometry.
(\textbf{m}) Schematic representation of the experimentally accessible regions of momentum space showing different
behaviors of the high-energy dispersion in different BZs immediately below 75~eV photon energy. Positive $k_x$ values
correspond to the experimental geometry approaching normal incidence. (\textbf{n}) Energy distribution curves taken at
the $\Gamma$ point from spectra shown in panels (\textbf{i}) and (\textbf{k}), showing a distinct bottom of the
renormalized band at about 0.5 eV.} \label{f:Maps}
\end{figure*}

The existence (or non-existence) of a high energy scale near 0.4~eV is of fundamental importance for the dressing of the
charge carriers in high-$T_\text{c}$ superconductors. This dressing may be related to the strange normal state
properties of these materials and possibly even to the mechanism of high-$T_\text{c}$ superconductivity. Furthermore,
``waterfalls'' have been detected for the first time in the angle resolved photoemission (ARPES) spectra of cuprates,
but never in any other correlated or uncorrelated material. Hence, the clarification of this phenomenon is also of great
importance for the ARPES method itself, which has now developed into one of the most powerful experimental methods in
solid state physics.

In this letter we report on the photon energy and momentum dependence of the ``waterfalls'' in high-$T_\text{c}$
superconducting cuprates. We observe strong differences in the shape of the single-particle excitation spectrum between
different BZs and its strong dependence on the excitation energy. This indicates that photoemission matrix elements
strongly influence the recorded spectral weight and that the reported values for a high energy scale, as well as the
respective physical models, may be incorrect.

\textit{Experiment}. In the current study we have used high quality single crystals of slightly overdoped
(Bi,\,Pb)$_2$Sr$_2$CaCu$_2$O$_{8+\delta}$ ($T_\text{c}=71$\,K), slightly underdoped
(Bi,\,Pb)$_2$Sr$_2$Ca$_{1-x}$Tb$_x$Cu$_2$O$_8$, optimally doped Bi-2212, and nearly optimally doped untwinned
YBa$_2$Cu$_3$O$_{6.85}$ ($T_\text{c}=92$\,K). The samples were cleaved \textit{in situ} in ultra-high vacuum
$\sim$\,$1\!\cdot\!10^{-10}$~mbar at room temperature and measured within 24 hours after cleavage at $\sim$\,20~K. ARPES
experiments were performed at the UE112-lowE PGMa beamline of the Berliner Elektronenspeicherring-Gesellschaft für
Synchrotron Strahlung m.b.H. (BESSY) with sub-meV energy resolution of the incident light. The spectra were acquired
using a \textit{Scienta R4000} electron analyzer with 30$^\circ$ acceptance angle and 1~meV energy resolution. The
overall energy and angular resolutions (including thermal broadening) were 10~meV and 0.2$^\circ$, respectively. In our
experimental geometry the analyzer slit is positioned at 45$^\circ$ to the synchrotron beam perpendicular to the
polarization direction of the incoming light.

\textit{Results}. Here we present several counter-examples which show that the ``waterfalls'' do not necessarily reveal
a ``new energy scale''. In Fig.\,\ref{f:Maps} we show several typical photoemission spectra of Bi-2212 along
high-symmetry directions (panels (a)\,--\,(d)) and the constant-energy maps at the Fermi level (panels (e) and (g)) and
at 380~meV below it for Bi(Pb)-2212 and Y-123 (panels (f) and (h)). As can be seen from comparison of panels (a) and
(b), presenting the spectra taken along equivalent cuts in momentum space in the 1st and 2nd BZ, the high-energy kinks
and \,\hbox{\waterfalls-}shaped ``waterfalls'' appear in the 1st BZ, while in the 2nd BZ neither of these features is
observed. Since both the electronic band structure and many body effects remain invariant under any translation by a
reciprocal lattice vector, the difference between these two images can come only from the photoemission matrix elements
which, as a rule, strongly depend on momentum and excitation energy \cite{BansilLindroos99,Borisenko01,Asensio03}.

In panels (i)\,--\,(l) we show the energy dependence of the spectra along the $(\pi,0)$ and $(\pi,\pi)$ directions in
the 2nd BZ taken with the photon energy close to the binding energy of the Cu\,3p level (75.1~eV), where the
photoionization cross section is modified by interchannel coupling of the direct photoemission process with an Auger
decay of the photoexcited Cu\,3p core hole \cite{DavisFeldkamp81}. Here we also observe an abrupt transition from
a~\hbox{\waterfalls-}shaped to a \hbox{\champagne-}shaped dispersion at $75\pm1$~eV photon energy. Panels (i) and (k)
show spectra below the transition that are to be compared with the equivalent spectra shown in panels (j) and (l) above
the transition energy. Using different experimental geometries, i.\,e. different sample positions relative to the
analyzer, we can access the 2nd BZ both at $k_x > 0$ (experimental geometry approaching normal incidence) and $k_x < 0$
(experimental geometry approaching grazing incidence). It is remarkable that we do not see a distinct transition neither
in the 1st, nor in the 2nd BZ at $k_x < 0$, as shown schematically in panel (m). At $\Gamma$ points marked by
``\kern1.2pt\waterfalls\kern.5pt'', the \hbox{\waterfalls-}shaped dispersion persists at all energies, while at the
$\Gamma$ point marked by ``\kern1.2pt\champagne\,'', a sharp transition from the~\hbox{\champagne-} to the
\hbox{\waterfalls-}like behavior is observed at 75~eV photon energy. In panel (n) we show energy distribution curves at
the $\Gamma$ point extracted from spectra (i) and (k), where a distinct band bottom at about 0.5 eV is observed.

However, the matrix elements can not explain all of the high-energy effects. As can be clearly seen both in Bi-2212
(panels (b), (d), (f), and (i)) and Y-123 (panel (h)), additional incoherent spectral weight is aggregated along the
bonding directions in the momentum space, i.\,e. $(2\pi n,k_y)$ and $(k_x,2\pi n)$, $n\in \mathbb{Z}$, persisting deeply
below the saddle-point of the conductance band (panel (d)) and forming a grid-like structure in the momentum space
(panel (h)). At the center of the 1st BZ this incoherent component is suppressed by matrix elements together with the
coherent part of the spectrum, forming the ``waterfalls'' (two long vertically dispersing tails seen in panels (a) and
(c)) and high-energy kinks.

In addition, we should mention that in our studies we have not detected any significant dependence of the high-energy
dispersion neither on doping nor on temperature.

\textit{Discussion}. Both photon energy dependence near 75~eV and the dependence on the BZ may be related to the
resonant enhancement of the Cu photoionization cross section at this energy \cite{Henke93}. In the Auger process which
resonates at $h\nu = 75$~eV with the normal photoemission process, near the threshold the core hole is produced by a
Cu~3p\,$\rightarrow$\,3d$_{x^2-y^2}$ transition. According to the dipole selection rules \cite{Fink94} this transition
is allowed for $\vec{E}$ vectors parallel to the surface of the sample (or to the Cu~3d$_{x^2-y^2}$ orbital) and
forbidden for $\vec{E}$ vectors perpendicular to the surface. Thus we would expect an enhancement of the Cu ionization
cross section above the resonance for those momenta (sample positions), for which the component of $\vec{E}$ parallel to
the surface is significant, i.\,e., for $k_x > 0$. This would mean that changing the photon energy from below to above
the resonance, small changes in the spectral weight should be expected for $k_x < 0$ and large changes should be
expected for $k_x > 0$, which is in good agreement with our experimental findings.

This suggests that in those BZs where we see for $h\nu < 75$ eV a \hbox{\champagne\kern.3pt-\kern.3pt}shaped dispersion,
we directly probe the spectral function of the renormalized band without significant distortion, while in the other BZs
and at higher excitation energies the \hbox{\waterfalls\kern.3pt-\kern.3pt}shaped ``waterfalls'' are produced by a
strong suppression of the spectral weight near the $\Gamma$ point due to matrix element effects \cite{MatrixElements}.
According to such interpretation, the peak seen near 0.5~eV in the energy distribution curves extracted from the
\hbox{\champagne-}shaped spectra (Fig. 1(n)) would be the bottom of the renormalized conduction band. It is interesting
to compare the observed band width with recent calculations of the band renormalization due to a coupling to the charge
carrier plasmon \cite{Markiewicz07}. There the renormalization factor $Z=0.5$ has been derived, corresponding to the
bandwidth narrowing by a factor of two, which is in reasonable agreement with the observed bottom of the conduction band
at 0.5 eV.

Finally, we discuss the vertical feature close to $(\pi, 0)$ which extends from $\sim$\,0.1 to 1.0~eV. Along the cut (d)
in Fig.~1, it has almost constant intensity below the saddle point of the conductance band, visible at a variety of
excitation energies. It is interesting that its distribution in momentum space is localized along the bonding directions
(Fig.~1(h)). In agreement with this, along the $(0, 0)$\,--\,$(2\pi, 0)$ cut (or equivalent) no feature is observed at
$(\pi, 0)$ neither at low ($h\nu=50$~eV) nor at high ($h\nu=100$~eV) photon energies (see e.\,g. Fig.~1(b)). But
surprisingly, in a similar cut taken with $h\nu=70$~eV (see Fig. 1(i)) we also see vertical features. They might stem
from the shadow bands, caused by the orthorhombic lattice distortions of Bi-2212, which are seen cutting the line ``a''
in Fig.~1(e) near the $(\pi, 0)$ point at an angle of $90^{\circ}$. Evidently there is a strong enhancement of the
shadow bands near the photon energy $h\nu=70$~eV, which is natural, because strong matrix element effects of the
spectral weight of these bands have been previously detected \cite{Mans06}.

Up to now there is no clear understanding of the source of the additional spectral weight along the bonding directions,
which manifests itself as the vertical feature at $(\pi,0)$ below the saddle point (Fig.~1(d) and (i)) and as the
``waterfalls'' at the $\Gamma$ point extending below the bottom of the conductance band in the energy range between 0.5
and 1.0~eV (Fig.~1(i) and (k)). Here we mention only that such an additional component is supported by recent optical
experiments \cite{HwangNicol06}. Evidently, this component, either incoherent or extrinsic, represents a new phenomenon
which deserves more systematic studies as a function of photon energy and momentum. Possible explanations can be related
with the disorder-localized in-gap states \cite{AlexandrovReynolds07}. The inelastic scattering of photoelectrons
\cite{Tougaard82} can be another option. On the other hand, the grid-like momentum distribution of this additional
spectral weight may hint at the presence of a one-dimensional structure \cite{ZhouScience99}. If so, then the
photoemission spectra consist of two components: one from the well studied two-dimensional metallic phase and another
from an underdoped one-dimensional phase. Such a scenario would be consistent with the ``checkerboard'' structure
observed by scanning tunneling spectroscopy in lightly hole-doped cuprates \cite{HanaguriNature04}.

\textit{Conclusions}. In the present study we have demonstrated that the recently observed ``waterfalls'' present in the
ARPES spectra of cuprates are to a large extent a result of the matrix element effects. This could indicate that the
postulated new high-energy scale in cuprates near 0.4~eV is not inherent in the single-particle spectral function of
these materials. At least we emphasize that due to matrix element effects this new energy scale can be hugely distorted,
complicating correct determination of the real dispersion, which is crucial for the high-$T_\text{c}$ superconductivity
problem, where understanding the nature of coupling requires the knowledge of very fine details of both the one-particle
and two-particle spectra \cite{Kordyuk06Fink07}. Still, we note that the present results do not doubt the
renormalization of the bare band at high energies ($\omega > 100$~meV) due to coupling of the charge carriers to spin
fluctuations, plasmons or other bosonic excitations \cite{Kordyuk06Fink07, KordyukBorisenko040506}.

\textit{Acknowledgements}. This~project~is~part~of~the~Forschergruppe~FOR538 and is supported by the DFG under Grants
No. KN393/4. The work in Lausanne was supported by the Swiss National Science Foundation and by the MaNEP. ARPES
experiments were performed at the UE112-lowE PGMa beamline of the Berliner Elektronenspeicherring-Gesellschaft für
Synchrotron Strahlung m.b.H. (BESSY). We thank R. Hübel for technical support.

\end{document}